\documentclass{article}
\usepackage{fleqn,espcrc2,amsmath,amssymb}
\usepackage[dvips]{graphicx}

\newcommand{\bsl}{\boldsymbol}

\begin{document}

\title{The doubly heavy baryons}

\author{I.M.Narodetskii and M.A.Trusov \\
{\itshape ITEP, Moscow, Russia}}

\begin{abstract}
We present the results for the masses of the doubly heavy baryons
$\Xi_{QQ'}$ and $\Omega_{Q'Q}$ where $Q,Q'=b,c$  obtained in the
framework of the simple approximation within the nonperturbative
string approach.
%The simple analytical results for
%dynamical masses of heavy and light quarks and eigenvalues of the
%effective QCD Hamiltonian are presented.
\end{abstract}

\maketitle

Doubly heavy baryons are baryons that contain two heavy quarks,
either $cc$, $bc$ or $bb$. Their existence is a natural
consequence of the quark model of hadrons, and it would be
surprising if they did not exist. In particular, data from the
BaBar and Belle collaborations at the SLAC and KEK B-factories
would be good places to look for doubly charmed baryons. Recently
the SELEX, the charm hadroproduction experiment at Fermilab,
reported a narrow state at $3519\pm 1$ MeV decaying in
$\Lambda_c^+K^-\pi^+$, consistent with the weak decay of the
doubly charged baryon $\Xi_{cc}^+$ \cite{Mattson02}. The candidate
is $6.3\sigma$ signal. Another candidate particle is related by
the replacement of a down by an up quark, so the mass difference
is expected to be similar to that of the proton and neutron.
SELEX, however, sees a mass difference 60 times larger
\cite{Cooper02}. Whether or not the states that SELEX reports turn
out to be the first observation of doubly charmed baryons,
studying their properties is important for a full understanding of
the strong interaction between quarks.

The purpose of this talk is to present the results of the
calculation \cite{NT01} of the masses
%and wave functions
of the doubly-heavy baryons obtained in a simple approximation
within the nonperturbative QCD (for a recent review see
\cite{Si02} and references therein). The starting point of the
approach is the Feynman-Schwinger representation for the Green
function of the three quarks propagating in the nonperturbatibe
confining background. The role of the time parameter along the
trajectory of each quark is played by the Fock-Schwinger proper
time. The proper and real times for each quark related via a new
quantity that eventually plays the role of the dynamical quark
mass. The final result is the derivation of the Effective
Hamiltonian, see Eq. (\ref{EH}) below. In contrast to the standard
approach of the constituent quark model the dynamical masses $m_i$
are no longer free parameters.  They are expressed in terms of the
running masses $m^{(0)}_i$ defined at the appropriate scale of
$\mu\sim 1$~GeV from the condition of the minimum of the baryon
mass $M_B$ as function of $m_i$: \begin{equation} \label{frac}
\frac{\partial M_B(m_i)}{\partial m_i}=0. \end{equation}
Technically, this has been done using the einbein (auxiliary
fields) approach, which is proven to be rather accurate in various
calculations for relativistic systems.

This method was already applied to study baryon Regge trajectories
\cite{FS91} and very recently for computation of magnetic moments
of light baryons \cite{KS00}. The essential point of this talk is
that it is very reasonable that the same method should also hold
for hadrons containing heavy quarks.
%In what follows we will
%concentrate on the masses of double heavy baryons.
As in \cite{KS00} we take as the universal parameter the QCD
string tension $\sigma$. We also include the perturbative Coulomb
interaction with the frozen coupling $\alpha_s(\text{1 GeV})=0.4$.

From experimental point of view, a detailed discussion of the
excited $QQ'q$ states is probably premature. Therefore we consider
the ground state baryons without radial and orbital excitations in
which case tensor and spin-orbit forces do not contribute
perturbatively. Then only the spin-spin interaction survives in
the perturbative approximation. The EH has the following form
\begin{equation}
\label{EH}
H=\sum\limits_{i=1}^3\left(\frac{m_i^{(0)2}}{2m_i}+
\frac{m_i}{2}\right)+H_0+V,
\end{equation}
where $H_0$ is the non-relativistic kinetic energy operator,
$m_i^{(0)}$ are the current quark masses and $m_i$ are the
dynamical quark masses to be found from (\ref{frac}), and $V$ is
the sum of the perturbative one gluon exchange potential $V_c$ and
the string potential $V_{\text{string}}$. The string potential has
been calculated in \cite{FS91} as the static energy of the three
heavy quarks: $V_{\text{string}}(\bsl{r}_1,\bsl{r}_2,
\bsl{r}_3)=\sigma R_{\text{min}}$, where $R_{\text{min}}$ is the
sum of the three distances $|\bsl{r}_i|$ from the string junction
point.

The baryon wave function depends on the three-body Jacobi
coordinates \begin{equation}\label{rho}
\bsl{\rho}_{ij}=\sqrt{\frac{\mu_{ij}}{\mu}}(\bsl{r}_i-\bsl{r}_j),
\end{equation}\begin{equation}\label{lambda}\bsl{\lambda}_{ij}=\sqrt{\frac{\mu_{ij,k}}{\mu}}
\left(\frac{m_i\bsl{r}_i+m_j\bsl{r}_j}{m_i+m_j}-\bsl{r}_k\right)\end{equation}
($i,j,k$ cyclic), where $\mu_{ij}$ and $\mu_{ij,k}$ are the
appropriate reduced masses
\begin{equation}\mu_{ij}=\frac{m_im_j}{m_i+m_j},~~
\mu_{ij,k}=\frac{(m_i+m_j)m_k}{m_i+m_j+m_k},\end{equation} and
$\mu$ is an arbitrary parameter with the dimension of mass which
drops off in the final expressions.

In terms of the Jacobi coordinates the kinetic energy operator
$H_0$ is written as
\begin{eqnarray} \label{H_0_jacobi} H_0&=& -\frac{1}{2\mu}
\left(\frac{\partial^2}{\partial\bsl{\rho}^2}
+\frac{\partial^2}{\partial\bsl{\lambda}^2}\right)\nonumber\\
&&=-\frac{1}{2\mu}\left( \frac{\partial^2}{\partial
R^2}+\frac{5}{R}\frac{\partial}{\partial R}+
\frac{K^2(\Omega)}{R^2}\right), \end{eqnarray} where $R$ is the
six-dimensional hyper-radius \begin{equation}
R^2=\bsl{\rho}^2+\bsl{\lambda}^2,\end{equation} and $K^2(\Omega)$
is angular momentum operator whose eigen functions (the
hyperspherical harmonics) are
\begin{equation}
\label{eigenfunctions} K^2(\Omega)Y_{[K]}=-K(K+4)Y_{[K]},
\end{equation}
with $K$ being the grand orbital momentum. In terms of $Y_{[K]}$
the wave function $\psi(\bsl{\rho},\bsl{\lambda})$ can be written
in a symbolical shorthand as
$$\psi(\bsl{\rho},\bsl{\lambda})=\sum\limits_K\psi_K(R)Y_{[K]}(\Omega).$$

In the hyper radial approximation  which we shall use below $K=0$
and $\psi=\psi(R)$. Note that the centrifugal potential in the
Schr\"odinger equation for the radial function
$\chi(R)=R^{5/2}\psi_K(R)$ with a given $K$ $$
\frac{(K+2)^2-1/4}{R^2}$$ is not zero even for $K=0$.

In terms of the $\bsl{\rho}$ and $\bsl{\lambda}$ the potential
$V_{\text{string}}(\bsl{r}_1,\bsl{r}_2, \bsl{r}_3)$ has rather
complicated structure. Let $\theta_{ijk}$ be the angle between the
line from quark $i$ to quark $j$ and that from quark $j$ to quark
$k$. If $\theta_{ijk}$ are all smaller than 120$^o$, then the
equilibrium junction position ${\bf Y}$ is \begin{equation} {\bf
Y}={\bf R}_{cm}+\alpha\bsl{\rho}+\beta\bsl{\lambda}, \label{torr}
\end{equation} where

\[
\begin{aligned}
\alpha&=\frac{1}{2}\sqrt{\frac{\mu}{\mu_{ij}}}\left(\frac{m_j-m_i}{m_i+m_j}-
\frac{1}{\sqrt{3}}\cdot \frac{4t+(3-t^2)\cot\chi}{1+t^2}\right),\\
\beta&=\frac{\sqrt{\mu\mu_{ij,k}}}{m_i+m_j}+
\sqrt{\frac{\mu}{3\mu_{ij}}}\cdot
\frac{\rho}{2\lambda\sin\chi}\cdot\frac{3-t^2}{1+t^2},
\end{aligned}
\]
where
\[
t=\frac{2\lambda\sin\chi+\sqrt{\dfrac{3\mu_{ij,k}}{\mu_{ij}}}\rho}
{2\lambda\cos\chi+\sqrt{\dfrac{3\mu_{ij,k}}{\mu_{ij}}}\cdot\dfrac{m_j-m_i}{m_i+m_j}\rho},
\]
and $\chi$ is the angle between $\bsl{\rho}$ and $\bsl{\lambda}$.
If $\theta_{ijk}$ is equal to or greater than 120$^o$, the lowest
energy configuration has the junction at the position of quark
$j$. It can be easily seen that dependence on $m_i$ in equations
(\ref{torr}) is apparent and $\bsl{Y}$ does not depend on quark
masses as it should be.

\begin{table}
\label{tab1} \caption{The constituent quark masses $m_i$ and the
ground state eigenvalues $E_0$ (in units of GeV) for the various
baryon states. } \vspace{2mm}

\begin{center}
\begin{tabular}{|c|c|c|c|c|}
\hline baryon & $m_1$ & $m_2$ & $m_3$ & $E_0$ \\ \hline (qqq) &
0.372 & 0.372 & 0.372 & 1.426 \\ \hline (qqs) & 0.377 & 0.377 &
0.415 & 1.398
\\ \hline (qss) & 0.381 & 0.420 & 0.420 & 1.370 \\ \hline (sss) & 0.424 &
0.424 & 0.424 & 1.343 \\ \hline (qqc) & 0.424 & 0.424 & 1.464 &
1.171
\\ \hline (qsc) & 0.427 & 0.465 & 1.467 & 1.146 \\ \hline (ssc) &
0.468 & 0.468 & 1.469 & 1.121 \\ \hline (qqb) & 0.446 & 0.446 &
4.819 & 1.085 \\ \hline (qsb) & 0.448 & 0.487 & 4.820 & 1.059 \\
\hline (ssb) & 0.490 & 0.490 & 4.821 & 1.033 \\ \hline (qcc) &
0.459 & 1.498 & 1.498 & 0.904 \\ \hline (scc) & 0.499 & 1.499 &
1.499 & 0.881 \\ \hline (qcb) & 0.477 & 1.524 & 4.834 & 0.783 \\
\hline (scb) & 0.517 & 1.525 & 4.834 & 0.759 \\ \hline (qbb) &
0.495 & 4.854 & 4.854 & 0.593 \\ \hline (sbb) & 0.534 & 4.855 &
4.855 & 0.570 \\ \hline
\end{tabular}
\end{center}
\end{table}

In what follows the string junction point is chosen as coinciding
with the center--of--mass coordinate. Accuracy of this
approximation that greatly simplifies the calculations was
discussed in \cite{FS91}. Averaging the interaction $V=V_c+
V_{\text{string}}$ over the six-dimensional sphere one obtains the
Schr\"odinger equation for $\chi(R)$
\begin{equation} \label{shr}
\frac{d^2\chi(R)}{dR^2}+2\mu\left[E_0+\frac{a}{R}-bR-\frac{15}{8\mu
R^2}\right]\chi(R)=0, \end{equation} where $E_0$ is the ground
state eigenvalue and
\begin{equation} \label{ab} a=\frac{2\alpha_s}{3}\cdot
\frac{16}{3\pi}\cdot\frac{1}{\sqrt{\mu}}\cdot \left(
\sum\limits_{i<j}\sqrt{\mu_{ij}}\right),\end{equation}\begin{equation}
b=\sigma\cdot\frac{32}{15\pi}\cdot\sqrt{\mu}\cdot\left(\sum\limits_{i<j}\frac{\sqrt{\mu_{ij,k}}}{m_k}\right),
\end{equation}

We use
%the same parameters as in Ref. \cite{KN00}:
$\sigma=0.15\text{~GeV}^2$, $\alpha_s=0.39$, $m^{(0)}_q=0.009$
GeV, $m^{(0)}_s=0.17$ GeV, $m^{(0)}_c=1.4$ GeV, and
$m^{(0)}_b=4.8$ GeV, slightly different from \cite{NT01}. We solve
Eq. (\ref{shr}) by the variational method introducing a simple
variational Ans\"atz \begin{equation}\label{gaussian}\chi(R)\sim
R^{5/2}e^{-\mu p^2R^2},\end{equation} where $p$ is the variational
parameter. Then the three-quark Hamiltonian admits explicit
solutions for the energy and the ground state eigenfunction:
$E\approx\min\limits_pE(p)$, where \begin{eqnarray}
E(p)&=&\langle\chi|H|\chi\rangle=
3p^2\nonumber\\&&-a\sqrt{\mu}\cdot\frac{3}{4}\sqrt{\frac{\pi}{2}}\cdot
p+\frac{b}{\sqrt{\mu}}\cdot\frac{15}{16}\frac{\sqrt{\frac{\pi}{2}}}{
p}. \end{eqnarray}

We first solve Eq. (\ref{frac}) for the  dynamical masses $m_i$
retaining only the string potential in the effective Hamiltonian
(\ref{EH}). This procedure is in agreement with the strategy
adopted in Ref. \cite{KS00}. Then we add the perturbative Coulomb
potential and solve Eq. (\ref{EH}) to obtain the ground state
eigenvalues $E_0$. The results for various baryons are given  in
Table 1. The dynamical values of light quark mass
$m_q\sim\sqrt{\sigma}\sim~400-500$ MeV ($q=u,d,s$) qualitatively
agree with the results of Ref. \cite{KN00} obtained from the
analysis of the heavy--light ground state mesons. For the heavy
quarks ($Q~=~c$ and $b$) the variation in the values of their
dynamical masses $m_Q$ is marginal. This is illustrated by the
simple analytical results for Qud baryons. These results were
obtained from the approximate solution of equation
\begin{equation} \frac{\partial E_0}{\partial p}=0\end{equation} in the form
of expansion in the small parameters
\begin{equation}\xi=\frac{\sqrt{\sigma}}{m_Q^{(0)}}~~ {\rm and}~~ \alpha_s.\end{equation}
Omitting the intermediate steps one has \begin{equation}
E_0=3\sqrt{\sigma}\left(\frac{6}{\pi}\right)^{1/4}\left(1+A\cdot\xi
-\frac{5}{3}B\cdot\alpha_s+\dots\right)\end{equation}
\begin{equation}
m_q=\sqrt{\sigma}\left(\frac{6}{\pi}\right)^{1/4}\left(1-A\cdot\xi+B\cdot\alpha_s+
\dots\right),\end{equation} \begin{equation}\label{m_Q}
m_Q=m_Q^{(0)}\left(1+{\cal O}(\xi^2,\alpha_s^2,
\alpha_s\xi)+\dots\right), \end{equation} \vspace{1mm}

\noindent where for the Gaussian variational Ans\"atz
(\ref{gaussian})\begin{equation}
A=\frac{\sqrt{2}-1}{2}\left(\frac{6}{\pi}\right)^{1/4}\approx
0.24,~ B=\frac{4+\sqrt{2}}{18}\sqrt{\frac{6}{\pi}}\approx
0.42.\end{equation} Note that the corrections of the first order
in $\xi$ and $\alpha_s$ are absent in the expression (\ref{m_Q})
for $m_Q$. Accuracy of this approximation is illustrated in Table
1 of Ref. \cite{NT02}.

\begin{table}
\caption{Masses of doubly heavy baryons}
\begin{center}
\begin{tabular}{|c|c|c|c|c|c|}
\hline State & this work &\cite{BDGNR94}& \cite{E97} & \cite{LRP95}& \cite{LO99}\\
\hline $\Xi\{qcc\}$ & 3.64 & 3.70 & 3.71 & 3.66 & 3.48\\
$\Omega\{scc\}$ & 3.82 & 3.80 & 3.76 & 3.74 & 3.58\\ \hline
$\Xi\{qcb\}$    & 6.93 & 6.99 & 6.95 & 7.04 & 6.82      \\
$\Omega\{scb\}$  & 7.10 & 7.07 & 7.05 & 7.09 & 6.92     \\ \hline
$\Xi\{qbb\}$    & 10.14 & 10.24 & 10.23 & 10.24 & 10.09
\\ $\Omega\{sbb\}$   & 10.31 & 10.30 & 10.32 & 10.37 & 10.19
\\ \hline
\end{tabular}
\end{center}
\end{table}

\begin{figure}
\label{fig} \caption{Mass of $\Xi_{cc}^+$ as a function of the
running $c$--quark mass for $\sigma=0.15\text{~GeV}^2$ and
$\sigma=0.17\text{~GeV}^2$. The masses are given in GeV. Bold
points refer to the case $m_c^{(0)}=1.4\text{~GeV}$}

\begin{center}
\includegraphics[width=80mm,
keepaspectratio=true]{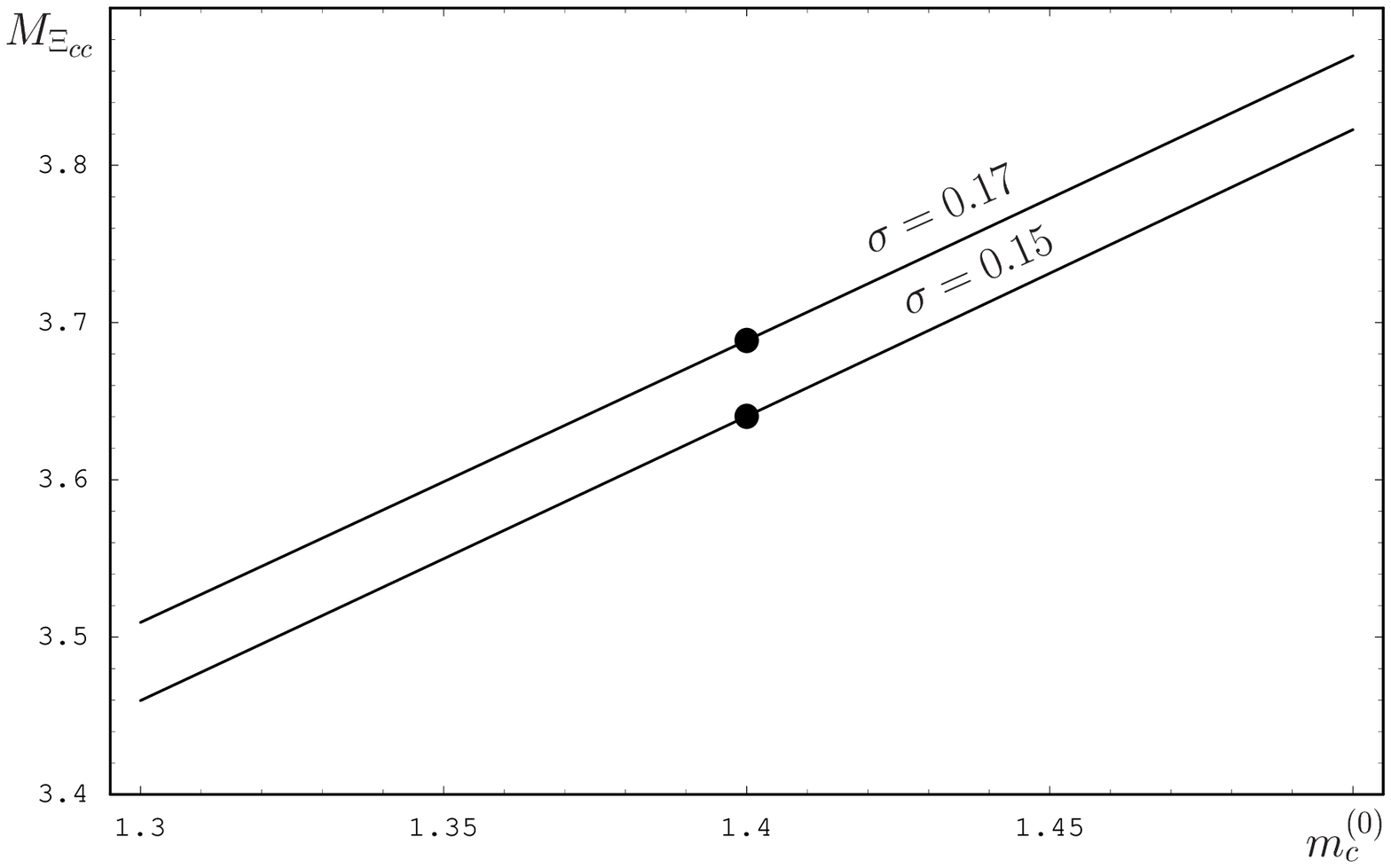}
\end{center}
\end{figure}

To calculate hadron masses we, as in Ref. \cite{FS91}, first
renormalize the string potential:  \begin{equation}
V_{\text{string}}\to
V_{\text{string}}+\sum\limits_iC_i,\end{equation} where the
constants $C_i$ take into account the residual self-energy (RSE)
of quarks \cite{Si01}. In what follows we adjust the RSE constants
$C_i$ to reproduce the center-of-gravity for baryons with a given
flavor. As a result we obtain $C_q=0.30\text{~GeV}$,
$C_s=0.15\text{~GeV}$, $C_c\sim C_b\sim 0.$

We keep these parameters fixed to calculate the masses given in
Table 2, namely the spin--averaged masses (computed without the
spin--spin term) of the lowest double heavy baryons. In this Table
we also compare our predictions with the results obtained using
the additive non--relativistic quark model with the power-law
potential \cite{BDGNR94}, relativistic quasipotential quark model
\cite{E97}, the Feynman-Hellmann theorem \cite{LRP95} and with the
predictions obtained in the  approximation of double heavy diquark
\cite{LO99}. The change of $\sigma$ to $0.17\text{~GeV}^2$
increases the mass of $\Xi_{cc}^+$ by $\sim 30$ MeV. The
perturbative spin-spin interaction introduces an additional shift
of the $\Xi_{cc}^+$ mass $\sim -20$ MeV. Note that the mass of
$\Xi_{cc}^+$ is rather sensitive to the value of the running
$c$-quark mass $m_c^{0}$, see Fig.1.

In conclusion, we have employed the general formalism for the
baryons, which is based on nonperturbative QCD and where the only
inputs are $\sigma$, $\alpha_s$ and two additive constants, $C_q$
and $C_s$, the residual self--energies of the light quarks. Using
this formalism we have also performed the calculations of the
spin--averaged masses of baryons with two heavy quarks. One can
see from Table 2 that our predictions are especially close to
those obtained in Ref. \cite{BDGNR94} using a variant of the
power--law potential adjusted to fit ground state baryons.

\section*{Acknowledgements}
This work was supported in part by RFBR grants \#\# 00-02-16363
and 00-15-96786.

\end{document}